\documentclass[12pt,a4paper]{article}
\usepackage{amsmath}
\usepackage{graphicx}
\usepackage{times}
\begin{document}
\begin{titlepage}
\title{Comment on  high-$p_T$ inclusive spectra measurements at the LHC }
\author{ S.M. Troshin, N.E. Tyurin\\[1ex]
\small  \it NRC ``Kurchatov Institute''--IHEP\\
\small  \it Protvino, 142281, Russian Federation}
\normalsize
\date{}
\maketitle

\begin{abstract}
The   high-$p_T$ inclusive spectra measured at the LHC demonstrate a non-perturbative $p_T^{-6}$ dependence. This  can be related
to observation of a nondecreasing with $p_T$ one-spin  asymmetries  at RHIC  questioning the fact of the spin degrees of freedom neglect at the LHC energies. 
\end{abstract}
\end{titlepage}
\setcounter{page}{2}

It is  well known  that the experimental observations on spin effects at various energies  have not found their ultimate explanation in QCD. The problems of perturbative QCD related to spin studies are well known since many years (cf. e.g \cite{nadol}).  Start of the LHC program has shifted the main interest of the physical community to another issues and spin is not actively  discussed nowadays at such high energies. The above is related to the LHC in the existing collider mode. Recent ideas on spin studies at the LHC energies can be found at \cite{after}.

Here we would like to  make a short comment on the possibility to study hyperon polarization in existing collider mode at the LHC. We think that such studies are not meaningless because high--$p_T$ hadron production allows  to interpret  the observed $p_T^{-6}$ inclusive spectra dependence as a nonperturbative effect \cite{koch,ter}. Indeed, an observation of high-$p_T$ tail in unpolarized inclusive processes deviating from the perturbative $p_T^{-4}$ dependence allows one to expect significant polarization of $\Lambda$-hyperons produced in the final state (cf. Fig. 1).
\begin{figure}[htb]
	\begin{center}
		\resizebox{6cm}{!}{\includegraphics*{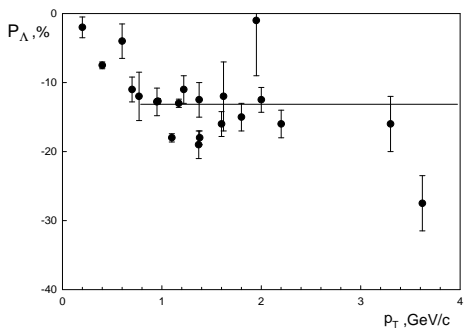}}
	\end{center}
	\caption{ $p_T$--
		dependence of the $\Lambda$-hyperon
		polarization.} \label{ts}
\end{figure}

The nonperturbative mechanism reproducing a $p_T^{-6}$-dependence of the inclusive spectra at high transverse momenta has been described in \cite{pepan} and has been revised recently with reference to the LHC spin measurements in \cite{spinlhc}. 

The important feature of this mechanism is an explanation of the flat nonzero single-spin asymmetries in the inclusive production.
It is based on the spin states filtering in case of the strange quark production responsible for a $\Lambda$-hyperon formation and is associated with the emission of the Goldstone bosons by the constituent quarks \cite{pepan} (Fig. 2).
\begin{figure}[h]
	\begin{center}
		\resizebox{6cm}{!}{\includegraphics*{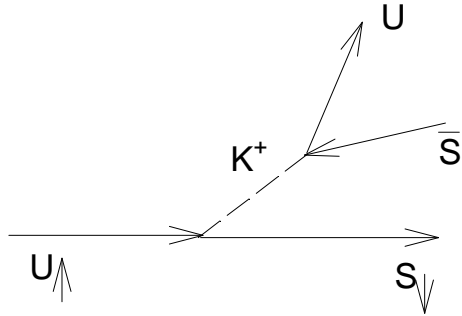}}
	\end{center}
	\caption{Transition of the spin-up constituent quark $U$ to the spin-down strange quark.
		\label{ts1}}
\end{figure}
 It leads to an energy-independent polarization value (cf. Fig. 1).
The arising final-state  hyperon polarization can be measured  due to a parity nonconservation in the hyperon's weak decay and there is no needs for polarized beam and/or polarized target.
It seems that, the most relevant experiments for such studies could be TOTEM and LHCb. 
 
This short comment on the possibility of spin studies at the LHC in collider mode was inspired by discussion with A. Penzo at the TOTEM meeting. We would also like to thank for discussions on the high--$p_T$ dependence of the unpolarized inclusive spectra Y. Kharlov, S.~Sadovsky and N. Topilskaya.


\small
\begin{thebibliography}{99}
\bibitem{nadol}
P.M. Nadolsky, S.M. Troshin, N.E. Tyurin, Int. J. Mod. Phys. A {\bf 9}, 2505 (1994).
\bibitem{after}
J.P. Lansberg at al. PoS PSTP2015, 042 (2015); arXiv: 1602.06857.
\bibitem{koch}
N. Kochelev, JETP Lett.  83, 527 (2006). 
\bibitem{ter}
T. Bhattacharyya, J. Cleymans, S. Mogliacci, A.S. Parvan, A.S. Sorin and O.V. Teryaev, 
arXiv: 1712.08334.
\bibitem{pepan}
S.M. Troshin, N.E. Tyurin, 
Phys. Rev. D {\bf 88}, 017502 (2013).
\bibitem{spinlhc}
S.M. Troshin, N.E. Tyurin,  arXiv: 1608.07940.
\end{thebibliography}
\end{document}